# Gold slows down the growth of helium bubble in iron

W. Hao and W. T. Geng[a]

*School of Materials Science & Engineering, University of Science & Technology Beijing, Beijing 100083, China*

January 20, 2011

We predict by first-principles calculations that Au have strong affinity to He in bcc Fe. The Au-Au bonding in the segregated Au layer at the He bubble surface is stronger than Fe-Fe and Au-Fe interactions; therefore this layer becomes an effective barrier to further He and slows down the bubble growth.

---

[a] To whom correspondence should be addressed. E-mail: geng@ustb.edu.cn

He bubbles in Fe-Cr alloys, which are considered to be promising structural materials used in fusion reactors, present a big challenge to maintaining the mechanical properties of these materials [1]. Accumulation of He atoms in Fe is nearly inevitable due to their strong self-trapping [2] and strong binding to vacancies [3] and grain boundaries [4]. On the other hand, the low diffusion barrier for interstitial He makes the growth of He clusters easier [5]. To combat He embrittlement, we need to impede both the growth and movement (or merge) of He bubbles. A recent work by Ono et al. demonstrated that in a Fe-9Cr% steel, Cr segregates to the He bubble surface and slows down the Brownian motion of small bubbles [6]. We have shown by first-principles density functional theory (DFT) calculations that this retarding effect is due to the increased charge density at the bubble/Fe interface upon Cr segregation, which makes it harder for further He atoms to passing through [7]. Inspired by this mechanism, we here propose a more powerful element to serve as the barrier to He. The candidate we select is Au, which has low surface energy than Fe and higher stability in low dimension (chain and monolayer) than any other metals [8]. Gold can form binary alloys with Fe [9], and its maximum solubility in bcc Fe can reach 1 at%, demonstrated by an experimental work in characterization of neutron-irradiated Fe-Au alloys [10].

First, we have examined if Au does segregate to the bubble surface. We argue that if Au is attracted to an individual He atom, it will also be attracted to a He bubble. We employed a (4×4×4) bcc Fe supercell. (Figure 1) We replaced two Fe atoms by one He and one Au, and investigated the interaction of He-Au by varying their distance. We find that He and Au attract each other strongly and binding energy, defined as the energy change of the supercell when the two move to neighboring sites from far apart, is as large as -0.56 eV. We then went on to evaluate the trapping energy for additional Au to the other seven nearest neighbor lattice sites of the He. The calculated trapping energies are listed in Table I. It is clear that all the nearest neighbors of He are favorable sites for Au to locate. An interesting feature shown in Table I is that the He-Au attraction does not reduce much as more Au are joining, especially for the first four atoms on one side of the He. This is a strong justification for using one substitutional He to represent a He bubble.

Our first-principle calculations were carried out using Vienna *ab initio* simulation package (VASP) [11] based on DFT. The projector augmented wave (PAW) method [12] was used to describe the electron-ion interaction and the exchange correlation between electrons was described by generalized gradient approximation (GGA) in the Perdew-Burke-Ernzerhof (PBE) form [13]. A cutoff energy of 500 eV was used for the plane-wave basis. The Monkhorst-Pack $k$-mesh was ($3\times3\times3$). The volume of the supercell was fixed but all the internal freedoms were fully relaxed. Lattice relaxation was continued until the forces on all the atoms were converged to less than $10^{-3}$ eV Å$^{-1}$. The lattice constant for bcc Fe, yielded by our GGA-PBE computation, was 2.83 Å.

With this computation model, we have calculated the diffusion barrier for a He atom in Fe to enter the Au cage and join the encapsulated He using the climbing nudged elastic band (CNEB) method [14]. The result is displayed in Figure 3. Strong self-trapping drives remote interstitial He atom into the Au cage to join with the caged He. The two He atoms favor a <100> dumbbell configuration in the Au cage. In the absence of Au, the barrier for a nearby interstitial He to join a substitutional He is essentially zero, in accordance with the fact that the diffusion barrier for interstitial He in bcc Fe is as small as 0.06 eV [Ref. 5]. Nonetheless, when covered by a layer of Au, the energy barrier for an interstitial He to join the substitutional He increases dramatically to 0.76 eV. As consequence, the growth of He bubble will be greatly slowed down.

Now, we turn to electronic structure analysis in order to understand the effect of segregated Au on the growth of He bubble. In Fig. 3 we plotted the calculated density of valence electrons around the He-8Au cluster. It is found that in this plane the Au-Au bonding is stronger than Au-Fe and Fe-Fe bonding, evidenced by a higher charge density in between Au-Au than in between Au-Fe and Fe-Fe. Since the 1s orbital of He does not hybridize with orbitals of any other atoms, it prefers in general to stay in a low-electron density position. As a result, it will be more difficult for an upcoming He to pass through the Au cage than an Fe cage.

The cohesive energy of elemental Au crystal (3.81 eV) is smaller than that of Fe (4.28 eV) [15]. However, the Au-Au bonds are stronger than Fe-Fe bonds at the Au-Fe interface. The reason is that the $5d$ states of Au are lower than $3d$ states of Fe, with respect of the Fermi energy. Therefore, the hybridization between Au and Fe is quite weak. This is clearly seen in the calculated local density of states (LDOS) of Au shown in Fig.4. With a He atom as nearest neighbor, the LDOS of Au is widened (blue to red). This is because a substitutional He yield a small free volume to Au and thus the latter is less compressed than otherwise. With the coming of additional Au atoms, LDOS of Au (green) is further dispersed as a result orbital hybridization between Au atoms. The enhanced bonding in low dimension of Au was attributed to the *s–d* competition caused by relativistic effects [16]. The contraction of *s* electron distribution for a 5th-row element is much more significant than that for 3rd- and 4th-row atoms. This contraction reduces the energy of 6*s* electron and increases its occupation number at the expense of 5*d* electrons. In the vicinity of a He atom, the spilling out of the 6*s* electron into free volume around He relieves its expansive Fermi pressure and leaves 5*d* contractive pressure uncompensated, thereby allowing a contraction of the nearest-neighbor distance with a strengthened interatomic bonding.

While Alloying Au into Fe increases strikingly the cost of materials, its magnificent retarding effect on bubble growth shed light on our design of structural materials for nuclear fusion reactors. Experimental scrutiny of our first-principles prediction is called for. Meanwhile, the search for cheaper atoms is warranted.

**Acknowledgments**

We are grateful to the support of the NSFC (Grant No. 50971029) and MOST (Grant No. 2009GB109004) of China, and NSFC-ANR (Grant No. 51061130558).

**Fig. 1 (Color Online) A gold cage encapsulating a He atom in bcc Fe. The numbers stand for the adding order of Au atoms.**

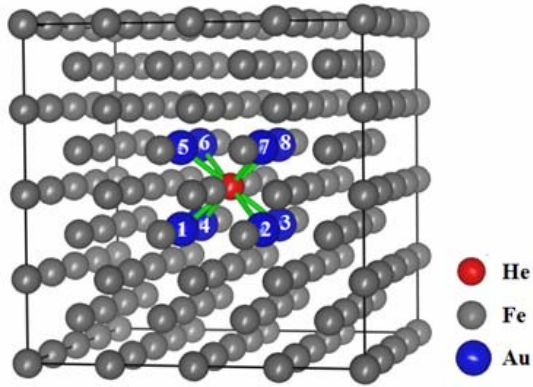

**Fig. 2 (Color online) Energy barrier for the upcoming interstitial He in bcc Fe to enter the Au cage (see Figure 1).**

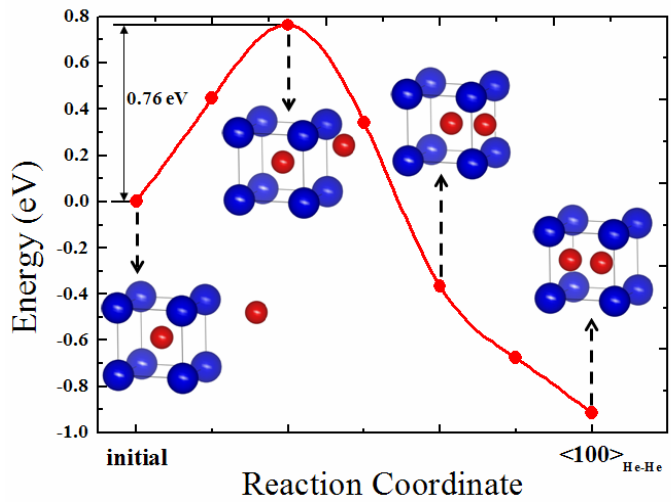

**Fig. 3 Distribution of valence electrons in the (001) plane containing Au1-Au4 of Fig. 1. Lines start from 0.025 $e$/a.u.$^3$ and increase successively by a factor of $10^{1/5}$.**

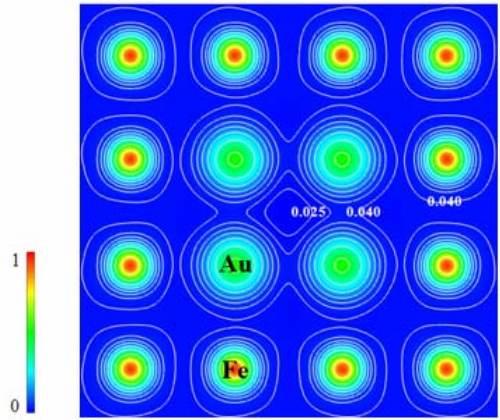

**Fig. 4 The calculated density of states (DOS) located at, respectively, an isolated Au, a sole Au atom near He, one of the 4 Au atoms near He, and a bulk Fe atom in bcc Fe. The Fermi energy is set to zero.**

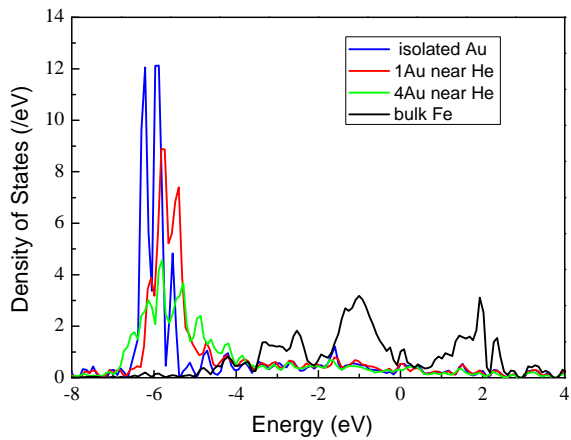

**Table I.** The calculated energy change when one Au atom is attracted from a perfect bcc Fe bulk environment to the vicinity of a substitutional He. The trapping process is in a sequential manner.

| $n$ | 1 | 2 | 3 | 4 | 5 | 6 | 7 | 8 |
|---|---|---|---|---|---|---|---|---|
| $\Delta E(n)$ /eV | -0.56 | -0.59 | -0.55 | -0.49 | -0.45 | -0.41 | -0.38 | -0.35 |